\documentclass[onecolumn]{aa}
\usepackage{natbib}
\bibpunct{(}{)}{;}{a}{}{,}
\usepackage{psfig}
\usepackage{epsfig,afterpage} 
\usepackage{graphicx}
\usepackage{pifont}
\usepackage{amssymb}
\usepackage{longtable}
\usepackage{array}
\usepackage{amsmath}
\usepackage{rotating}
\usepackage{multirow}
\usepackage{longtable}
\usepackage{lscape}

\begin{document}

\title {The Wilson--Bappu Effect: a tool to determine stellar distances.}
\thanks{Observations collected at ESO, La Silla}
  \subtitle{ }

\author{G. Pace\inst{1,2} \and L. Pasquini\inst{2} \and S.Ortolani\inst{3}}

\offprints{L. Pasquini, \email lpasquin@eso.org}

\institute{ Dipartimento di Astronomia, Universit\`a di Trieste, 
            Via G.B. Tiepolo 11, 34131 Trieste, Italy \\
       \and
            European Southern Observatory, 
            Karl Schwarzschild Strasse 2, 85748, 
            Garching bei M\"{u}nchen, Germany \\ 
       \and
            Dipartimento di Astronomia, Universit\`a di Padova, 
            Vicolo dell'Osservatorio 5, 35122 Padova, Italy \\
            }
\date{Received July 15, 2002; accepted July 16, 2002}

\abstract{
\citet{orig} have shown the existence of a remarkable correlation 
between the width of the emission in the core of the K line of
Ca $\!{\rm II}$\  and 
the absolute visual magnitude of late--type stars.

Here we present  a new calibration of the Wilson--Bappu effect
based on a sample of 119 nearby stars. We  use, for the first time, width 
measurements based on high resolution and high signal to  noise ratio CCD 
spectra and absolute visual magnitudes from the \textit{Hipparcos}\  database.

Our primary goal is to investigate the possibility of using the Wilson--Bappu 
effect to determine accurate distances to single stars and groups.

The result of our calibration fitting of the Wilson--Bappu relationship
is  $M_V=33.2-18.0\cdot \log W_0$, 
and the determination seems free of systematic effects.
The root mean square error of the fitting
is 0.6 magnitudes. This error is mostly accounted for by measurement 
errors and intrinsic variability of $W_0$, but in addition
a possible dependence on the metallicity is found, which becomes clearly 
noticeable for metallicities  below [Fe/H]$\sim$-0.4. This 
detection is possible because in our sample [Fe/H] ranges from -1.5 to 0.4.

The Wilson--Bappu effect can be used 
confidently for all metallicities not lower than $\sim$-0.4, including the 
LMC. While it  does not provide  accurate distances to single stars, 
it is a useful tool to determine accurate distances to 
 clusters and aggregates, where a sufficient number of stars can be observed. 

We apply the Wilson--Bappu effect to published data 
of the open cluster M67; the retrieved distance modulus is of 9.65 magnitude,
in very good  agreement with the best distance estimations for this 
cluster, based on main sequence fitting.  
      }
\titlerunning{Wilson--Bappu effect}

\maketitle

\section{Introduction}
The K line of  Ca $\!{\rm II}$\ is the deepest and widest 
absorption line in the spectra of the late type stars, and
its core shows the characteristic double reversal profile: an emission
 with a central self--absorption \citep[see e.g.][]{luca88}.

Since the discovery by \citet{orig} of the existence of a linear relationship
between the logarithm of the width of the Ca $\!{\rm II}$\ emission 
($W_0$) and the stellar absolute
visual magnitude (the so called Wilson--Bappu effect ), 
several  calibrations of this effect have been attempted.
However the Wilson--Bappu relationship (WBR) has only seldom 
been used to determine stellar distances to single stars or aggregates. 
 
The reliability of past calibrations of the WBR
has been limited by the lack of two crucial elements:
\begin{enumerate}
\item Precise and accurate parallaxes for nearby stars;
\item High quality data and objective method of measuring $W_0$.
\end{enumerate}
The first point has been overcome by \citet{wallerstein},
who produced a new calibration by  exploiting the \textit{Hipparcos}
\ database.
This work however was hampered by the fact that it was still  based on 
$W_0$ measurements 
provided by \citet{w67,w76}, performed on photographic  plates.
For our calibration we have used $W_0$ measurements performed on new, 
high quality CCD spectra,
so that also the second point has been, for the first time, properly 
accounted for.

A new, reliable WBR determination is especially interesting since
new detectors and state of the art spectrographs can now produce 
excellent Ca $\!{\rm II}$\ data even for stars
in stellar clusters and associations as distant as several kiloparsecs.
For these clusters and associations, we could therefore apply the 
WBR to derive their distance.
This possible application is the main ground of our effort to retrieve a 
reliable calibration for the WBR. 

\section{Data sample and observations}

The full sample for which spectra have been collected (data shown in Table
\ref{tabglob}) consists of 152 stars, but the present study is limited to 
stars 
with relative parallax errors smaller than $10\%$. We have also excluded 
from the original sample known multiple systems.
After this trimming, the final sample includes 119 stars.

All the stars but the Sun are included in the \textit{Hipparcos}\ 
catalogue, from which trigonometric parallaxes and visual magnitudes have
been taken.

The absolute visual magnitude of the Sun has been taken from 
\citet{hayes}
The sample has been selected in order to span a wide range of luminosities
(from $M_V \simeq  -5$ to $M_V \simeq  9$).

The observations were obtained between 
November 1988  and September 
1996, at ESO, La Silla, with the Coud\'e Echelle Spectrometer,
at the focus of the Coud\'e Auxiliary Telescope.
The  resolution is R=60000 and the S/N ratio ranges from $\simeq  30$ to 
$\simeq  100$ at the bottom of the line (for more details about the
first spectra see \citet{luca92}).

For $\sim$ 30 stars, multiple spectra were taken, and for
the Sun 9 spectra are available.

Spectral types and metallicities were obtained from the \citet{cayrel} 
catalogue. For the stars not included in this catalogue, spectral
types are taken from the \textit{Hipparcos}\ database. The projected rotation 
velocities are from \citet{gl00}.
  
\section{Measurements and calibration}
\label{measandcal}
$W_0$ was measured manually on all of the spectra.
We have also developed an IDL macro which performs a multi Gaussian 
fitting to the K--line double--reversal profile, but it has not been used 
for the final computation because  it has proven quite 
fragile for the correct determination of the broad 
line absorption profile. We are working on improving the method,
and we may report further progress elsewhere. 

$W_0$  has been computed as the difference in wavelength between 
the two points taken at the intensity equal to the average between those
of the K1 minimum and K2 peak on either side of the emission 
profile (see Figure \ref{goodspec}).

We found that the definition of $W_0$ which we have adopted correlates 
 better with $M_V$ than two other widths also measured: 
\begin{list}{$\cdot$}{\itemindent -15pt \itemsep -3pt }
\item the difference in wavelength between K1 minima; 
\item the width taken at the intensity equal to 
the average  between those of the K1 minima and K2 maxima.
\end{list}

No correction for instrumental broadening was applied to $W_0$. The reason
is explained in Sect.\ref{insteff}.

We notice that our definition of $W_0$ differs slightly from others 
used in the literature. 
\citet{orig} define $W$ as the difference in wavelength between the red edge
and the violet edge of the emission profile. They apply to the measured value
in the spectra a linear correction: $W_0= W - 15$ km $\cdot$ s$^{-1}$. 
\citet{w59} kept the original definition but applied a revised correction:
18 km $\cdot$ s$^{-1}$ instead of 15.
\citet{lutz70} introduced a new definition of $W_0$, defining it as the width 
at half of the maximum of the emission profile.
Lutz's  definition is very similar to ours with the exception that we found
the our definition easier to use in case of difficult spectra and more robust 
(see below).

$W_0$ measurements are subject, of course, to measurement errors. For each
star we have computed an accuracy qualifier, $\Delta W_0$, in the following 
way. For the stars  with two or more spectra available we derived it as the 
half of the
difference between the largest and smallest measured widths. For stars 
with only one available spectrum, we measured $W_0$ with multiple methods,
and took the difference between the extrema of the measurements.
Then we added quadratically to this value (which in some cases was 0) 
that of the Sun (0.030 \AA), whose $W_0$ measurements variations are  
supposed to be caused only by the intrinsic variation of
the line width and by the limit of the resolution power of the spectrograph.
We have used $\Delta W_0$ as an estimate of the mean error in $W_0$.
The standard error of $\log W_0$ (where $W_0$ is 
in km $\cdot$ s$^{-1}$) for each star, is retrieved applying the propagation of the mean 
error.

Independent of the signal to noise ratio, 
for some stars it is intrinsically more difficult to measure $W_0$.
This is because their spectra show asymmetric self absorption, 
either produced by  interstellar lines or by blueshifted winds or cosmic 
rays. Inactive, low luminosity stars will typically
show shallow reversals, which are more difficult to measure.

\begin{figure*}
\centering
\includegraphics{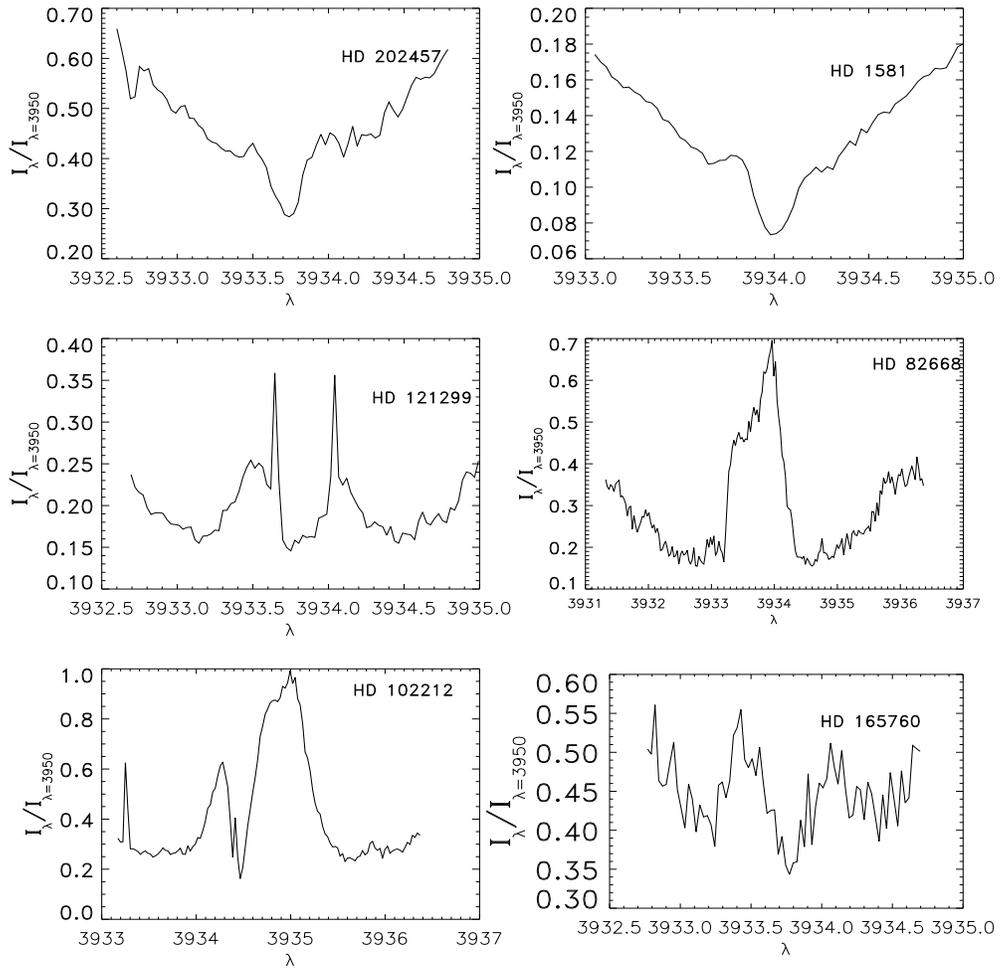}
\caption{Doubtful examples of Ca $\!{\rm II}$\ K line profiles. The spectra 
are affected by cosmic rays, may be blended with interstellar 
absorption, which strongly influence the emission profile observed in 
HD 82668, or, as in the case of HD 102212, show a blueshifted wind. In other
cases, such as in  HD 1581, the emission is very weak.}
\label{multifigure}
\end{figure*}

\begin{figure}
\resizebox{\hsize}{!}{\includegraphics{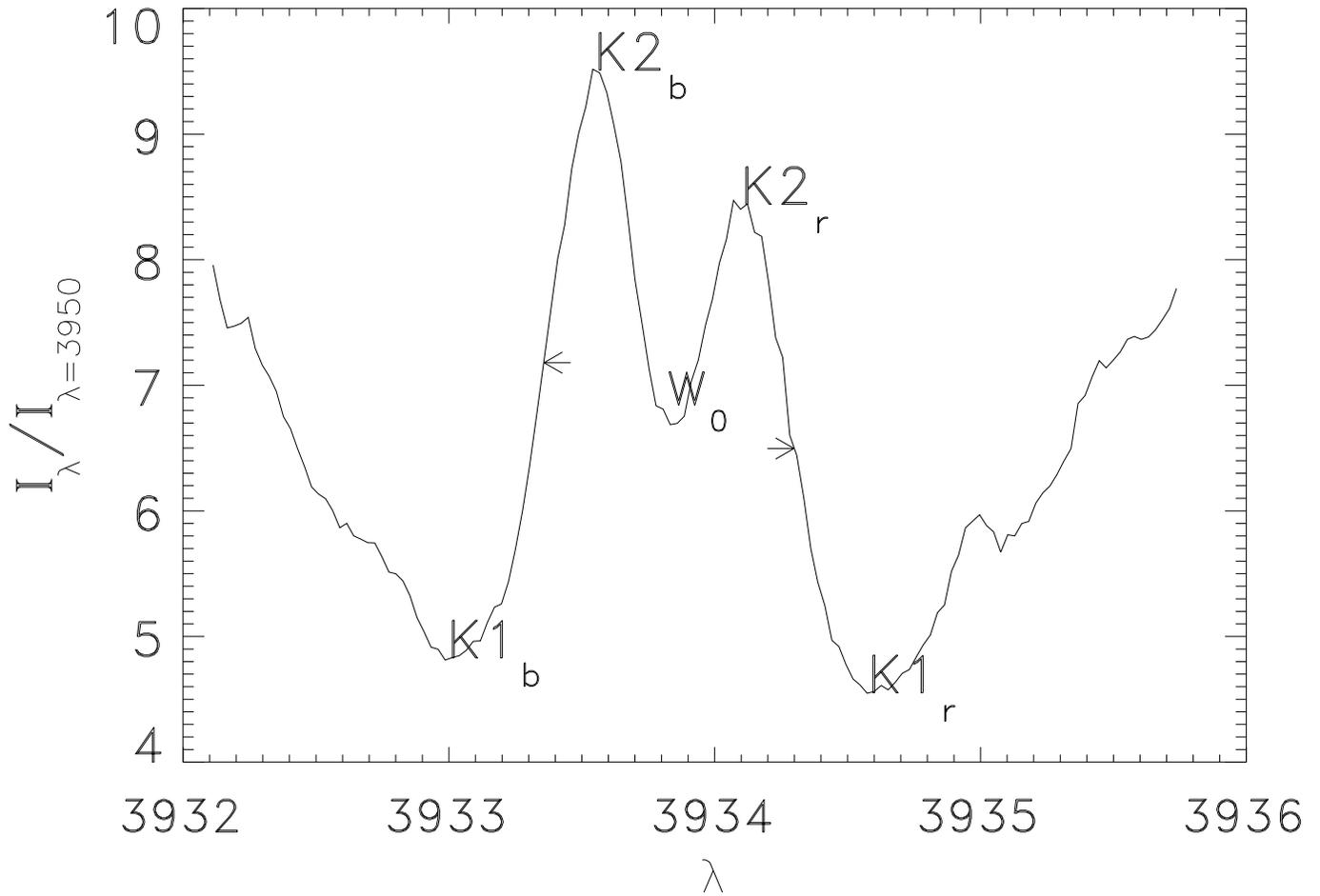}}
\caption{Spectrum of HD 4128. Most of the spectra of our sample have a 
comparable quality.}
\label{goodspec}
\end{figure}

Some of the most doubtful examples and difficult cases are presented in Figure
\ref{multifigure}, in order to make the reader acquainted 
with the spectra and the possible error sources. 
In some cases, when strong blending was present or the
profile was highly asymmetric, we have measured $W_0$ by doubling
the value measured for the ``clean'' half of the line.
Anyway, the majority of the spectra we dealt with were as good as the one 
showed  in Figure \ref{goodspec}.

We have also computed the standard error on the absolute magnitude 
of each star. This error has two components: the mean error on the apparent 
 visual magnitude 
(which is, in most of the cases, negligible) and the error given by the 
uncertainty in the parallax, which has to be computed via the propagation 
of the error. 

The fit of the WBR was performed by means of the IDL routine 
``fitexy'', which implements the algorithm described in \citet{numrec}.
The algorithm fits a straight line to a set of data points by 
taking into accounts errors on both coordinates.

We repeated the solution by rejecting stars not passing $3\sigma$--, 
$2.5\sigma$-- or $2\sigma$--criteria. The results so obtained are presented 
in Table \ref{sclip}

It is fundamental to note that, independent of the different sigma clipping 
 criterion used, the solutions found are extremely stable, giving the 
same fit to within $1\sigma$.

We  adopt in the following: 
$M_V=33.2-18.0\cdot\log{W_0}$ 
($W_0$ is in km $\cdot$ s$^{-1}$) obtained rejecting HD 63077 and HD 211998,
with a standard deviation: $\sigma_{WBR}=0.6$ mag.
The two rejected stars are the most metal poor of the sample,
which we will argue 
in Sect. \ref{meteff} is the most likely cause of their large 
residuals.

Hereafter we indicate with $M_V(K)$ the value of $M_V$ derived
for a single star from its $W_0$ via the WBR.
In Figure \ref{cal.all} the $\log W_0$ vs $M_V$ diagram is shown, with the 
error bars representing standard errors in both 
coordinates. The calibration line retrieved is also plotted.

\section{Comparison with other results}
\label{comp}
In order to evaluate the external robustness of the 
relationship,  a comparison with other independent investigations is 
fundamental. 
Many attempts in the past 40 years have been made to calibrate the WBR
based on ground--based trigonometric parallaxes
\citep{w59,hodge,lutz75,gl78}. \citet{wallerstein}  (hereafter WMPG) used 
a very large sample of stars with  \textit{Hipparcos}\ 
parallaxes and relative  measurement errors smaller than 20\%. As already 
stated, their work is based on the width measurements available in the 
literature, mostly from \citet{w67,w76}, wich suffer from low
spectral accuracy.\newline \noindent
We note here that such measurements are based on the 
definition of $W$  given in \citet{orig}, corrected for instrumental 
broadening as in \citet{w59} (see Sect. \ref{measandcal}).
So the quantity $W_0$ which they adopt is not exactly the same as ours, 
although the two quantities are expected to be strongly correlated.

WMPG used a linear least squares fitting both not weighted 
and weighted only in absolute magnitude with $\frac{\epsilon_{\pi}}{\pi}$.
As WMPG advised, using weighted least squares means giving more
weight to the lowest part of the diagram, containing the dwarfs that
are, on the average, much closer, and therefore with smaller measurement
errors on parallaxes.
Using the weight for both the coordinates, as we did, does not produce to
the same effect, because at the same time the dwarfs have also smaller
$W_0$, and so larger relative measurement errors, i.e. larger 
standard errors for  $\log W_0$.
In Figure \ref{confr} we show the comparison of our calibration and the
weighted calibration of WMPG.

We think that this comparison is a strong test 
of the reliability of the WBR. In fact,
with the only exception of having used \textit{Hipparcos}\ parallaxes, the two
samples are
independent: less than the $40\%$ of the stars we used in the computation
are common to WMPG. 
Different stars means different distributions
in magnitudes, metallicities and effective temperatures. Furthermore, 
different spectrographs were 
used, different measurements (even based on different definitions of 
$W_0$) were performed as well as a different analysis.
Taken at the face values, our calibration and WMPG's one
are in good agreement for all stars brighter than $M_V \sim$2, 
while for dwarfs, the discrepancy is stronger, reaching a difference of 1.5 
magnitudes at $M_V \sim 9$. 

In order to investigate how much
of this discrepancy is  due to the differences in $W_0$ measurements,
we compared, for the 64 stars common to the two data sets, the $W_0$
measurements of WMPG and ours. 
Actually 20 of these stars are among the 33 not used in 
our WBR computation, but for the present comparison this is irrelevant.
The  two sets of measurements show, as expected, a very strong linear 
correlation: the slope of the $W_{0_{our}}$ vs $W_{0_{WMPG}}$ linear 
fitting is very much closed to unity: 1.003, with an intercept of 
-5.34 km $\cdot$ s$^{-1}$ (see Figure \ref{cfrmeasures}).

If we subtract 5.34 km $\cdot$ s$^{-1}$ to our $W_0$ measurements, we obtain a data 
set homogeneous to that of 
WMPG, and performing the fitting with the new values gives
following WBR: 
$M_V=29.7-16.3\cdot \log W_0$.
This result matches very well that of WMPG, as it can be seen from Figure 
\ref{confr}.
We conlcude that the reliability of the WBR is excellent, and that the 
only reason
for the discrepancy between the calibrations is the difference in the 
definition  of $W_0$. 

On the other hand we notice that care is needed in 
measuring $W_0$: its definition, and possibly
the resolution of the spectra used shoud be the 
same as those of the calibration adopted.

The fact that the difference between our and WMPG's measurements is
about 5 km $\cdot$ s$^{-1}$, quite similar to  the projected slit width for R=60000,
could suggest that the instrumental profile should indeed be linearly
subtracted by
our $W_0$ measurements to obtain an-instrument free calibration.

We do not believe that this is the case, because:
\begin{enumerate}{\itemindent -15pt \itemsep -3pt }
\item Other studies \citep{lutz70} have shown as the case of linear 
subtraction of instrumental profile is not the best choice.
\item Our observations of one star (HD 36069) taken at different resolutions
to estimate this effect show  that the variations observed in $W_0$
measurements are not better accounted for by a linear sibtraction of the 
instrumental profile, as shown from the measurements in Table \ref{resol}.
\end{enumerate}

For sake of completeness, we remind that the WBR is also valid for 
the k--line of the Mg $\!{\rm II}$, and that the best calibration to date 
is the one of \citet{cassatella}, which also uses
\textit{Hipparcos}\ data and  IUE spectra. Their result 
is: $M_V=34.56-16.75\cdot \log W_0$

\section{Is the WBR a good distance indicator?}

The spread around the WBR
is still too large  to  consider it as a reliable distance indicator for
single stars. The question we are now going to investigate 
in this section is if the WBR 
is suitable to determine the distance of clusters of stars. 
A necessary condition which such clusters have to satisfy is, of course,
that for a sufficient number of members, a high quality spectrum, showing
a clear double reversal profile of the K--line, is available.

The possibility of using the WBR to determine accurate cluster distances is 
strictly related to the causes of the scatter:
whether or not it is due to entirely random errors or
systematic effects.
Among the possible causes of scatter, we mention:
\begin{list}{$\cdot$}{\itemindent -15pt \itemsep -3pt }
\item random effects:
  \begin{list}{-}{\itemindent -15pt \itemsep -3pt }
  \item measurement errors,
  \item cyclic variation in the chromospheric activity,
  \item variability of some of the stars of the sample;
  \end{list}

\item systematic effects:
  \begin{list}{-}{\itemindent -15pt \itemsep -3pt }
  \item reddening,
  \item instrumental effects,
  \item Lutz--Kelker effect (hereafter LKE),
  \item hidden parameters, i.e. parameters other than $W_0$ and $M_V$ on which
       the WBR could depend.
  \end{list}

\end{list}

\citet{white} observed the chromospheric emission of the K line
of the sun during a whole solar cycle. They found a maximum variation
of $\log W_0$ of about 0.05 during such a period. If we assume that most
of the stars are affected by a variation of $\log W_0$ of the same order of
magnitude, the amount of scatter introduced by the cyclic variation of the
chromospheric activity would represent a relevant fraction of the spread
observed in the data. Nevertheless, this variability cannot fully explain
the observed root mean square error of the WBR fitting. With typical
uncertainities in $W_0$ due to measurement errors and natural variations of
the stellar line width of about 0.036 \AA (cf. Table 1), this error, for stars
with intermediate widths, say $W_0$= 0.8 \AA, accounts for
about 0.35 mag of $\sigma_{WBR}$ = 0.6 mag. Therefore, it is necessary to 
investigate further reasons of uncertainty in the determination of the WBR.
 
Among possible causes of biases we should consider
reddening, the LKE (\citet{lutz73}, hereafter LKP), instrumental 
effects and the presence of multiple systems.

\subsection{Lutz--Kelker effect}

The LKE is the bias due to the fact that 
a symmetric error interval 
$[\pi-\sigma_{\pi},\pi+\sigma_{\pi}]$ 
around the  estimated parallax $\pi$, does not correspond to a symmetric error
interval in distances around $\frac{1}{\pi}$. The inner spherical corona
centred in the Sun having radii $\frac{1}{\pi+\sigma_{\pi}}$ and 
$\frac{1}{\pi}$, has a volume smaller than the outer spherical corona. So, 
assuming a homogeneous space density for the stars, we expect that, for a 
fixed measured parallax, stars having a true distance greater than 
$\frac{1}{\pi}$, i.e. those
in the outer corona, will outnumber the stars having a distance smaller than 
$\frac{1}{\pi}$. There is therefore a systematic trend to underestimate 
distances. The correction which has to be applied to each star, has been 
calculated in LKP. It depends only on the relative 
error $\frac{\sigma_{\pi}}{\pi}$.

Our sample has been selected to include only stars with 
($\frac{\sigma_{\pi}}{\pi} \leq 0.1$). Furthermore, out of the
119 stars, only 7 have 
$\frac{\sigma_{\pi}}{\pi}$ exceeding 0.075. For these values the LKE
 is negligible compared with other errors involved: 0.06 mag for 
$\frac{\sigma_{\pi}}{\pi} = 0.075$ and 0.11 mag for
$\frac{\sigma_{\pi}}{\pi} =0.1$ (See Table 1 in LKP)

\subsection{Reddening}
We have assumed that all stars have zero reddening. This assumption 
is justified by the fact that the sample stars are all within $\sim$200 
parsec.
\newline \noindent
The most distant star, HD 43455, has a distance of 205 pc, and it is the only 
one for
which we were not able to find out a secure upper limit to the reddening.
\newline \noindent
HD 78647 has a distance of 176 pc, and a galactic latitude 
lower than 7.6$^o$, so we can get a rough estimation of its reddening on the 
Neckel \& Klare's maps \citep{neckel}. For this star, $A_V$ does not 
exceed $\sim 0.1$ mag.
\newline \noindent
The remaining stars  are within 107 pc. According to \citet{sfeir} (see 
their Figure 2) the upper limit of the equivalent width of the D2 
Na $\!{\rm I}$ line for such a distance is 200 m\AA. From this quantity we can
get the Hydrogen column density \citep{welsh}:
N(H $\!{\rm I}$)$\sim2\cdot 10^{20}$, which yields a colour excess:
E$_{B-V}\sim 0.03$, or an upper limit for $A_V$ of about 0.1 mag.
\newline \noindent
Furthermore, 100 of the 119 stars in this sample, 
are within 75 parsec, so they are in the so called Local Bubble 
\citep[see e.g.][]{sfeir},
and they are not affected by detectable extinction.

\subsection{Multiple systems}
The presence of unrecognised multiple systems in the sample,
gives rise to a systematic underestimation of $M_V(K)$, 
because we could associate the width of the K line emission of 
one component to the magnitude of the whole system.
To avoid this effect, we checked all the objects of our sample on the SIMBAD
database and we have excluded all the known multiple systems.

\subsection{Instrumental effects}
\label{insteff}

The measured $W_0$ is likely larger than the intrisic one because of 
the broadening introduced by the spectrograph. The larger the projected 
slit  width is the stronger the intrumental broadening will be.
We have shown in Sect. \ref{comp}, by means of data in Table \ref{resol}, 
that a linear correction for instrumental broadening (i.e. subtracting
the projected slit width from $W_0$) would not be appropriate. Similar 
results were found by \citet{lutz70}, who concluded that a quadratical
correction should be used.
To minimize this effect, our calibration is based on high
resolution spectra (the projected slit width is about 0.066 \AA or 5 km $\cdot$ s$^{-1}$), 
and appling a quadratical correction even to the smallest $W_0$ 
value (that of HD 42581, 0.30 \AA) we would obtain: 
$W_0-W_{0_{corrected}}=W_0-\sqrt{W_0^2-(0.066)^2}=0.0074$\AA, 
well below its estimated measurement error, i.e. $\Delta W_0=0.02$\AA.
For larger values of $W_0$, $W_0-W_{0_{corrected}}$ is even smaller.
Hence, the quadratical correction is negligible for all stars in our sample.
%If we apply it to all stars and we compute back the WBR, the result changes
%very slightly:
%$M_V~=33.1 -17.9 \cdot \log W_0$, with the two same outlier data points.
We believe that the quadratical correction is more appropriate  
than the linear one, and it should be applyied when dealing with low
resolution spectra, but it is not certain that such a small adjustment 
would represent
a real improvement when dealing with data of resolution comparable to that 
used in this work.

\subsection{Other causes}
\label{meteff}
Many authors searched for additional parameters on which the WBR could depend,
finding contradictory results.
\citet{gl78} proposed a corrected  WBR with a term for the intensity of the 
emission that WMPG  have rejected. 

\begin{figure}[!h]
\centering
\resizebox{\hsize}{!}{\includegraphics{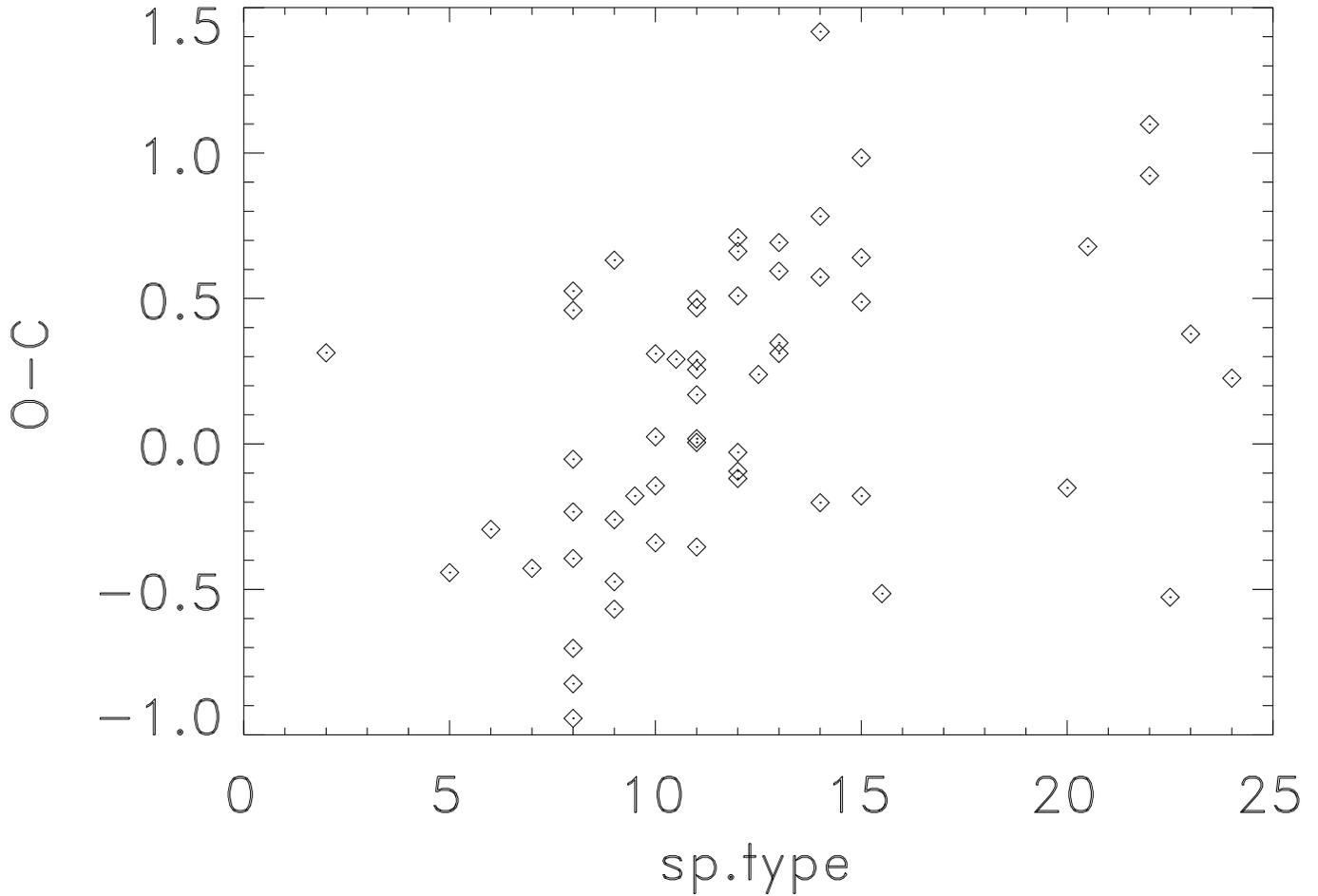}}
\caption{The spectral type diagram vs $O-C$. The spectral types are indexed 
as in \citet{parsons}: 0 is for G0 stars, 1 for G1 and so on. Only luminous 
stars  are plotted (no {\rm IV} and {\rm V} luminosity classes).}
\label{scartsp}
\end{figure}

\citet{parsons}, analysing the calibration of WMPG, 
suggested a trend for high luminosity stars that our data seem not to 
confirm:
he suggested that $O-C$ (i.e. the difference between the absolute magnitude
from \textit{Hipparcos}\ parallax and the one retrieved by means of the WBR)
increases with increasing $T_{eff}$ for spectral 
types earlier than $\sim$ K3, while the opposite is true for the other stars.
He also concludes that this trend gets stronger for brighter stars. 
According to Figure \ref{scartsp}, while we can draw no  conclusions for 
late type stars, our data seem to suggest a trend  opposite to that proposed 
by \citet{parsons} for spectral types earlier than $\sim$ K3.

The most obvious hidden parameter to search for is  projected rotational 
velocity. High rotational velocity can influence $W_0$ in several ways,  
 either because fast rotating stars will tend to be more active
\citep[see e.g.][]{cutispoto}, or because the 
width of the line core may be modified by the higher rotational velocity
\citep[see e.g.][]{luca89}. 
We have 53 stars for which $V\cdot \sin i$\ is  available, and none are
really fast rotators, only for one object $V\cdot\sin i$ exceeds
 10 km $\cdot$ s$^{-1}$. 
Our conclusion is that, among slow rotators, there is hardly any 
dependence of the residuals on $V\cdot\sin i$: we find a correlation 
coefficient of 0.12
 
We have finally  searched for a dependence of the  $O-C$ on metallicity. 
Such a dependence can also be expected, considering that
in stars having  lower abundances the core of the line may sample different 
layers of the atmosphere. 
 
\begin{figure*}[!ht]
\centering
\includegraphics{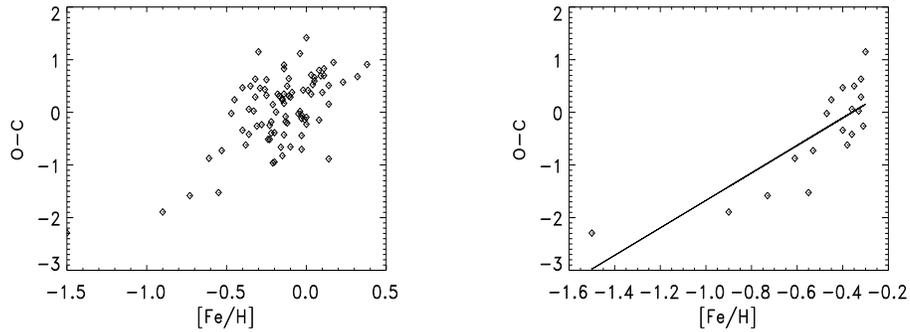}
\caption{The $O-C$ vs [Fe/H] diagrams both for all stars of the sample with 
available metallicities (on the left) and  for metal poor stars only.
On the latter is also shown the retrieved regression line, which has a slope
as high as 2.61.}
\label{metsc}
\end{figure*}

\begin{figure}
\resizebox{\hsize}{!}{\includegraphics{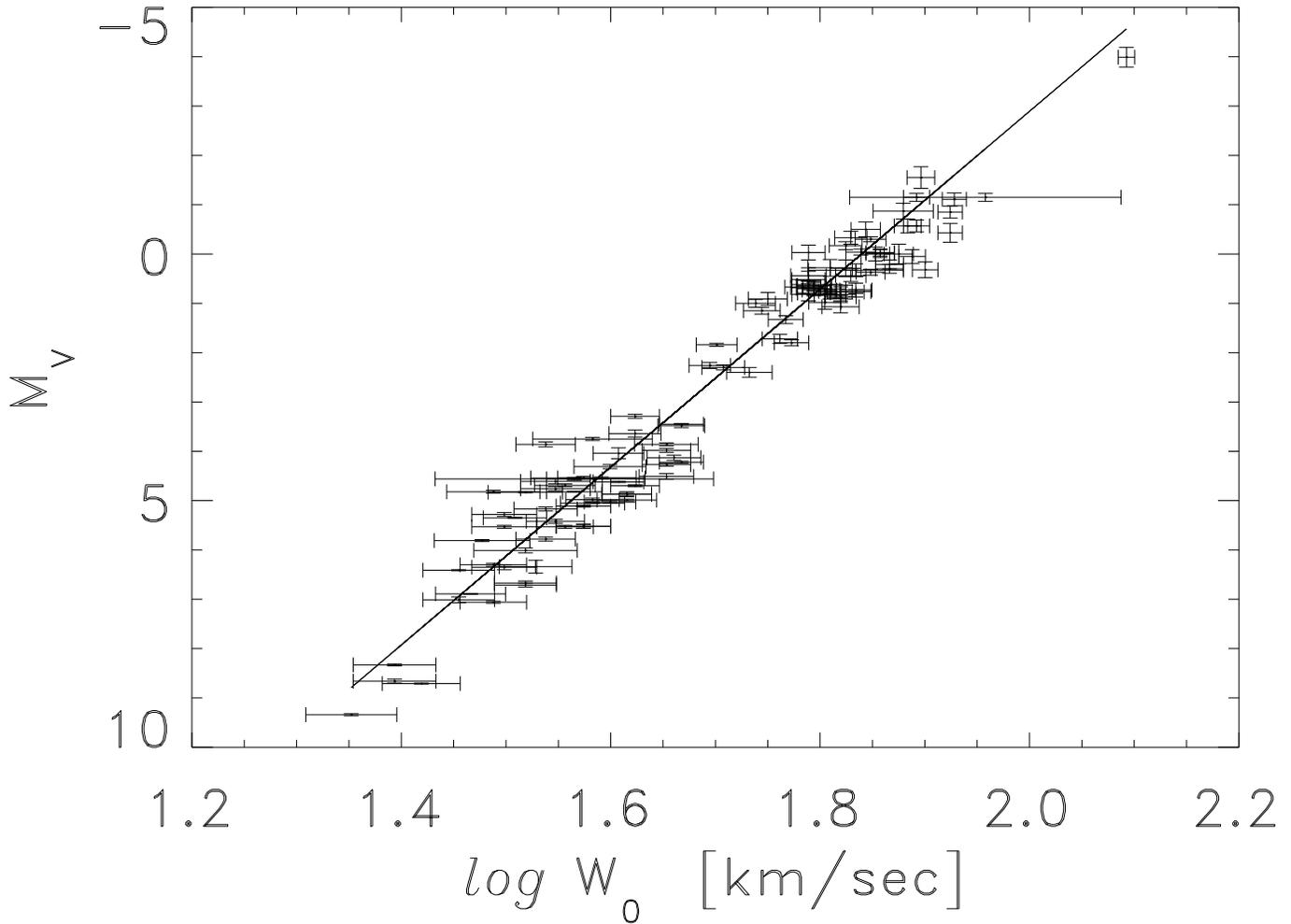}}
\caption{Our calibration of the Wilson--Bappu Effect:
$M_V~=~33.2~-~18.0~\cdot~\log W_0$.
This calibration is the $3\sigma$ criterion one in Table \ref{sclip}
(HD 63077 and HD 211998 are not used). The error bars represent standard 
errors in both the coordinates. }
\label{cal.all}
\end{figure}

\begin{figure}
\resizebox{\hsize}{!}{\includegraphics{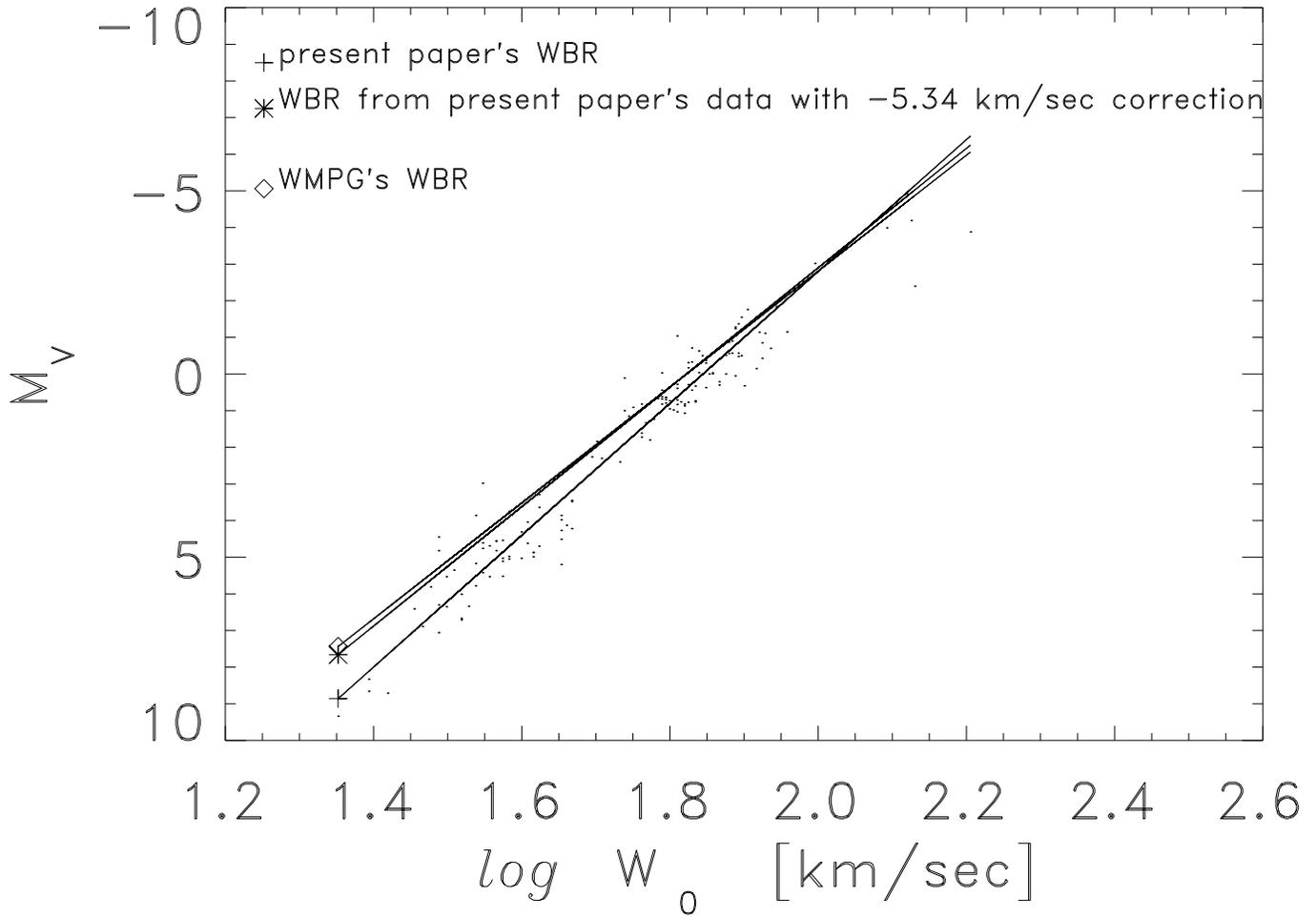}}
\caption{Comparison between the following calibrations:
the present paper's one  ($M_V~=33.2~-18.0~\cdot~\log~W_0$),
the one we obtained after subtracting 5.34 km $\cdot$ s$^{-1}$ to all $W_0$ measurements
and that of WMPG (the weighted one: 
$M_V~=~28.83~-~15.82~\cdot~\log~W_0$).
The points refer to our data, including stars not used in the calibration
because of the uncertainty in the parallax
($\frac{\sigma_{\pi}}{\pi}>0.1$) or because they are binary or multiple 
systems.}
\label{confr}
\end{figure}

\begin{figure}
\resizebox{\hsize}{!}{\includegraphics{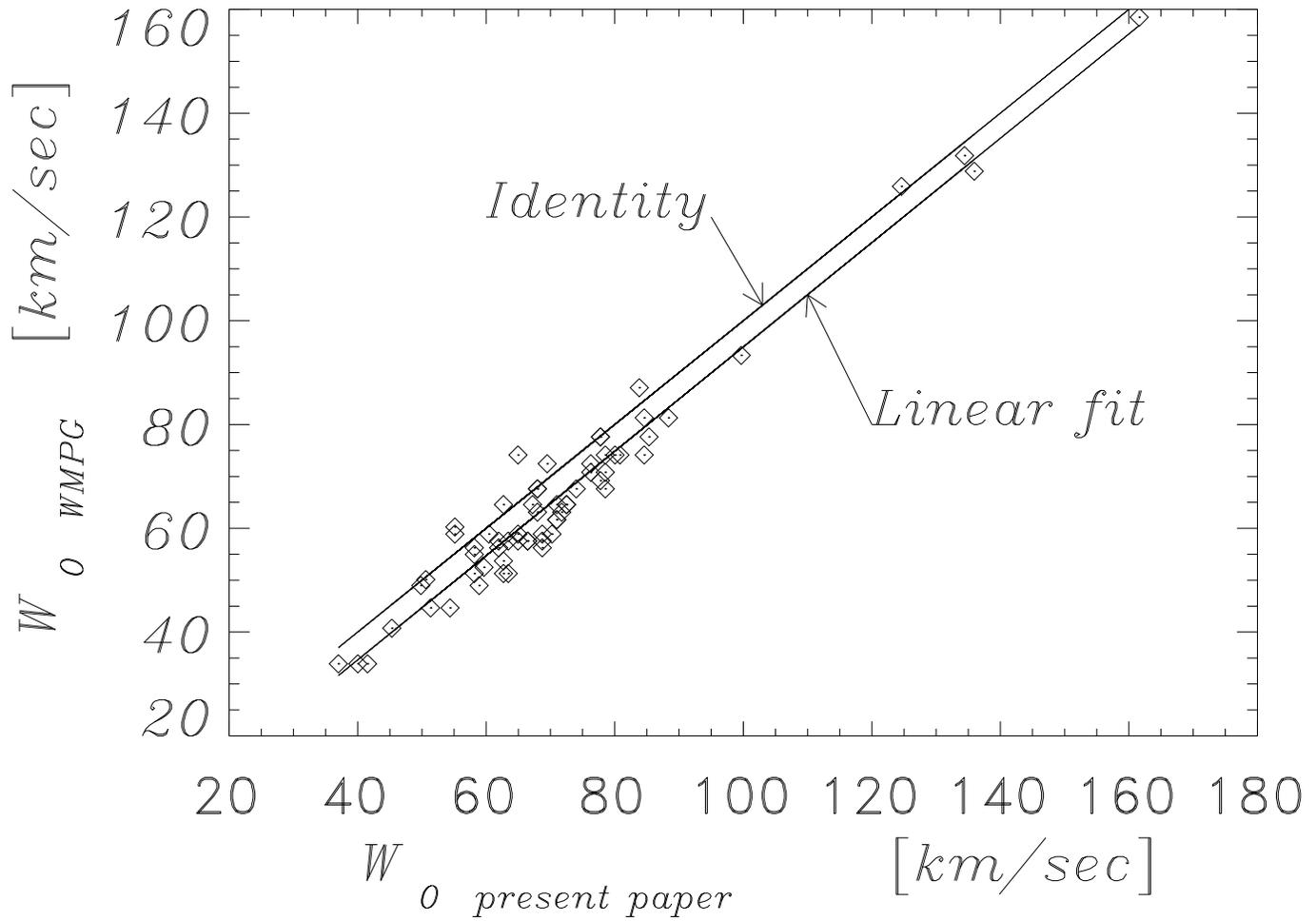}}
\caption{Comparison between our measurements of $W_0$ and those used by WMPG.}
\label{cfrmeasures}
\end{figure}

In particular we have checked
whether the WBR is still  valid for very metal poor stars.
Figure \ref{metsc}  shows two $O-C$ vs [Fe/H] diagrams: the one on the 
left refers to  all the 
stars with available metallicities, in the other diagram only stars with 
[Fe/H]$<-0.3$ are plotted.
A weak but not negligible dependence of the WBR on metallicity 
does exist, and it gets much stronger for metal poor stars. 
The correlation coefficient is 0.64, and it becomes 0.82 when the 19 
 most metal poor stars are considered, as shown on the right panel of
figure \ref{metsc}.

19 stars are too few to obtain any firm quantitative conclusions.
In particular, the $O-C$ vs [Fe/H] relationship, to which they would point out
(the straight line in right panel of Figure \ref{metsc}),
should be further investigated by means of a richer sample.
The existence of such a relationship for metal poor
stars has been independently suggested by \citet{dupree95}, who studied  
53 metal poor giants, none of which is in our sample.

We think that the WBR 
should be applied very carefully
to very metal poor  stars (e.g. stars more metal poor 
than [Fe/H]$\sim$-0.4) 
and that further metal poor calibrators should be observed before applying 
it to very metal poor clusters.

\section{Application to M67}

After deriving the WBR, and showing that the scatter is mostly due to
random errors, we have the opportunity to test 
it on a group of stars belonging to a well studied open cluster. 
M67 Ca $\!{\rm II}$\  spectra were published by \citet{dupree99} 
(See Figure 2, 3 and 4 therein) for 15 stars 
on the RGB and clump region, and they are suitable for our analysis
of the WBR. 
Andrea Dupree kindly provided us with all the spectra in digital form.

Since M67 has been extensively studied, 
the retrieved distance modulus can be compared with values obtained
from other authors.
\citet{carraro} provide a detailed study of M67. They derive, 
on the basis of the Colour Magnitude Diagram, $9.55 \le (m-M)_V \le 9.65$ mag.

\citet{montgomery} performed a photometric survey of the central region of
M67. They compared their photometry with two theoretical isochrones 
to retrieve distance modulus and age for M67.
From $V, B-V$ CMDs, they found $(m-M)_V$=9.60 
for both isochrones (but different ages were found), they have also used a 
$V,V-I$ CMD, giving $(m-M)_V$=9.85.
\citet{dinescu} found $9.7 \le (m-M)_V \le 9.8$ mag, obtained by letting
$E_{B-V}$ varying between its upper (0.06 mag) and lower (0.03 mag) limits.
Their isochrones were constructed using model atmospheres with new 
opacities.
In \citet{montgomery} other results from the literature are reported,
ranging from  9.55 to  9.61.

In summary, all distance modulus determinations for M67 are in the range 
$9.55 \le (m-M)_V \le 9.85$ mag.

M67 is a solar metallicity cluster, so we do not need to take care of the
metallicity effect which may affect the WBR.
The $W_0$ measurements were performed in the same way as for the 
calibration stars, and the results are given  in Table \ref{m67}.
Out of the 15 stars of the original sample we have selected a subsample
of 10, which suitable spectra were available, either for quality 
or clearness of the core reversal. 
In fact some  of the spectra do not show a clear unambiguously 
recognisable double reversal feature, so that the measurement is unreliable.
We did not use the stars with the following Sanders ID numbers
\citep{sanders} : 258, 989, 1074, 1316, 1279.
Even among the 10 selected stars 
some show a clearer profile than others, and for four of them the 
measurements were more uncertain (of the order of 0.1 \AA) 
and they have been flagged with an asterisk in Table \ref{m67}.

We have to consider that M67 spectra were acquired for other purposes, and in 
particular they have lower resolution and lower S/N ratio than
the typical calibration spectra, so we 
expect a standard error on the single measurement higher than the 
$\sigma_{WBR}$ derived above.

In Table \ref{m67}  the distance modulus determinations for the single
stars retrieved by means of the WBR are given. They range from 8.1 to 10.9 
mag. The mean value is $\sim$9.7
In spite of the poorer quality of the spectra, all the deviation can be 
explained on the base of the intrinsic spread around the WBR.

For sake of accuracy we have also taken into account the effect of the 
difference in resolution between the calibration spectra and the M67 
observations (5 and 11 km $\cdot$ s$^{-1}$ respectively). 
A simple, quadratic correction 
for the difference between the two projected slit widths 
is applied in the sixth column of Table \ref{m67}.
The correction does not change the result in an appreciable way.

When considering all stars a simple mean gives (M-m)=9.61 mag; 
which becomes 9.65 when discarding the 4 most uncertain measurements.

We expect that the standard error in our determination of the distance of M 67
would be about: $\frac{\sigma_{WBR}}{\sqrt{6}}\sim0.3$ if we used 6 
spectra of quality similar to 
those used for our calibration ($\sigma_{M67}^{\prime}\sim0.2$ if we had 10 
spectra of the same quality).

Trying to push further this application would definitely represent a gross
over interpretation of the data, however we find it extremely interesting and 
encouraging that a simple application, using published data, 
can provide a distance modulus in the range between 9.5 and 9.8, 
in excellent agreement with completely independent measurements, such as 
those obtained with main sequence model fitting. 
%-----------------
\section{Conclusions} 

We have shown that the coupling of CCD high resolution, high S/N ratio data 
with the use of the \textit{Hipparcos}\ parallaxes allows a good determination
of the WBR. 
The root mean square error 
found around this relationship (0.6 magnitude) is not 
good enough 
to determine 
accurate distances to single stars, but it can be used to infer 
accurate distances of clusters or groups, provided that they are not too
metal poor.
This is possible because the uncertainties 
in the relationship are mostly due to random  errors
(measurements, cycles) and not from systematic effects. 
This implies that once one has observed a sufficient number
of stars,  $n$, the distance modulus standard error can be reduced to about 
$ {0.6}$mag$/{\sqrt{n}}$.
Its extension to metal poor objects (e.g. stars with Fe/H $<$-0.4) 
would require extra care to fully evaluate the impact of low metallicity on 
the relationship.
When using our WBR in photometric parallax determinations, the resolution 
used should be comparable (within a factor $\sim 3$)
to that of the calibration (R=60000), to avoid large corrections,
 and care has to be exercised in measuring  $W_0$, following the proper
calibration definition.

\begin{acknowledgements}
We are greatly indebted to P. Bristow and N. Bastian for their careful 
reading of the manuscript.
We thank the referee, Elena Schilbach, for very valuable comments and 
suggestions, which  improved considerably the quality of this paper.
Special thanks to A. Dupree, who kindly provided us with the M67 spectra.
\end{acknowledgements}

\bibliographystyle{aa}

\begin{thebibliography}{}

\bibitem[Carraro et al., 1996]{carraro} 
  Carraro, G., Girardi, L., Bressan A., \& Chiosi, C. 1996, A\&A 305, 849

\bibitem[Cassatella et al., 2001]{cassatella}
Cassatella, A., Altamore, A., Badiali, M., \& Cardini, D. 2001, A\&A 374. 1085 

\bibitem[Cayrel de Strobel et al., 1997]{cayrel} 
  Cayrel de Strobel, G., Soubiran, C., Friel, E.D., Ralite, N., \& Francois P. 1997, A\&AS 124, 229

\bibitem[Cutispoto et al., 2002]{cutispoto} 
  Cutispoto, G., Pastori, L., Pasquini, L., de Medeiros, J.R., Tagliaferri, G., \& Andersen, J. 2002, A\&A 384, 491

\bibitem[Dinescu et al., 1995]{dinescu} 
  Dinescu, D. I., Demarque, P., Guenther, D. B., \& Pinsonneault, M. H. 1995, AJ109, 2090

\bibitem[Dupree \& Smith, 1995]{dupree95}
  Dupree, A. K., \& Smith, G. H. 1995, AJ 110, 405 

\bibitem[Dupree et al., 1999]{dupree99} 
  Dupree, A. K., Whitney, B. A., \& Pasquini, L. 1999, ApJ 520, 751

\bibitem[Hayes, 1985]{hayes}
  Hayes, D.S. 1985, IAU Symp. 111, 347

\bibitem[Gl\d{e}bocki \& Stawikowski, 1978]{gl78} 
  Gl\d{e}bocki, R., \& Stawikowski, A. 1978, A\&A 68, 69

\bibitem[Gl\d{e}bocki et al., 2000]{gl00} 
  Gl\d{e}bocki, R., Gnacinski, P., \& Stawikowski, A. 2000, AcA 50, 509

\bibitem[Hodge \& Wallerstein, 1966]{hodge} 
  Hodge, P.W., \& Wallerstein, G. 1966, PASP 78, 411 

\bibitem[Lutz, 1970]{lutz70} 
  Lutz, T.E., 1970, AJ 75, 1007

\bibitem[Lutz \& Kelker, 1973]{lutz73} 
  Lutz, T.E., \& Kelker, D.H. 1973, PASP 85, 573

\bibitem[Lutz \& Kelker, 1975]{lutz75} 
  Lutz, T.E., \& Kelker D.H. 1975, PASP 87, 617

\bibitem[Montgomery et al., 1993]{montgomery} 
  Montgomery, K. A., Marschall, L. A., \& Janes, K. A. 1993, AJ 106, 181

\bibitem[Neckel et al., 1980]{neckel} 
  Neckel, Th.,  Klare, G., \&  Sarcander, M. 1980, A\&AS 42, 251

\bibitem[Parsons, 2001]{parsons} 
  Parsons, S.B. 2001, PASP 113, 188

\bibitem[Pasquini et al., 1988]{luca88} 
  Pasquini, L., Pallavicini, R., \& Pakull, M. 1988, A\&A 191, 253

\bibitem[Pasquini et al., 1989]{luca89} 
  Pasquini, L., Pallavicini, R., \& Dravins, D. 1989, A\&A 213, 261

\bibitem[Pasquini, 1992]{luca92} 
  Pasquini, L. 1992, A\&A 266, 347

\bibitem[Press et al., 1989]{numrec} 
  Press, W.H., Teukolsky, S.A., Vetterling, V.T., \& Flanney, B.P. 1992, Numerical Recipies (Cambridge Univ. Press), 660

\bibitem[Sanders, 1977]{sanders} 
  Sanders, W. L. 1977, A\&AS 27, 89

\bibitem[Sfeir et al., 1999]{sfeir}
  Sfeir, D. M., Lallement, R., Crifo, F., \& Welsh, B. Y. 1999, A\&A 346, 785 

\bibitem[Wallerstein et al., 1999]{wallerstein} 
  Wallerstein, G., Machado--Pelaez, L., \& Gonzalez, G. 1999, PASP 111, 335

\bibitem[Welsh et al., 1994]{welsh} 
  Welsh, B. Y., Craig, N., Vedder, P. W., \& Vallerga, J. V.1994, ApJ, 437, 638

\bibitem[White \& Livingston, 1981]{white} 
  White, O.R., \& Livingston, W. C. 1981, ApJ 249, 798

\bibitem[Wilson \& Bappu, 1957]{orig} 
  Wilson, O.C., \& Bappu, M.K.V. 1957, ApJ 125, 661 

\bibitem[Wilson, 1959]{w59} 
  Wilson, O.C. 1959, ApJ 130, 499

\bibitem[Wilson, 1967]{w67} 
  Wilson, O.C. 1967, PASP 79, 46

\bibitem[Wilson, 1976]{w76} 
  Wilson, O.C. 1976, ApJ 205, 823

\end{thebibliography}

%%%%%%%%%%%%%%%%%#############################
\clearpage

\begin{longtable}{c c c c c c c c c c }
\caption{\label{tabglob}Our data sample. 
The full sample consists of 152 stars, 33 of which have
not been used in the present study, because of  their high  measurement 
error in the parallax
($\frac{\sigma_{\pi}}{\pi}>0.1$) or because they are multiple 
systems. We have flagged their HD identificators with an asterisk.
Column 1: HD ID number of the star.
Column 2: $W_0$ in \AA. 
Column 3: accuracy qualifier for $W_0$, again in \AA. Its mean value is 0.036.
Column 4: absolute visual magnitude using \textit{Hipparcos}\ parallaxes.
For the Sun we used the value given by \citet{hayes}.
Column 5: spectral type from Cayrel De Strobel Catalogue. When the spectral 
type is not available from this catalogue the data is taken from the 
\textit{Hipparcos}\ database, and is flagged with a dagger.
Column 6: mean metallicity, when available from Cayrel De Strobel Catalogue.
Column 7: projected rotation velocity, when available from \citet{gl00}.} \\
\hline
\hline
 Star &$W_0$ [\AA] &$\Delta W_0$ [\AA] &$M_V$&Sp.Type&[Fe/H]&$V\cdot\sin i$\\
\hline
\endfirsthead
\multicolumn{10}{l} {Table \ref{tabglob} cont.}\\
\multicolumn{10}{c}{}\\
\hline
\hline
 Star &$W_0$ [\AA] &$\Delta W_0$ [\AA] &$M_V$&Sp.Type&[Fe/H]&$V\cdot\sin i$\\ 
\hline
\endhead
\hline
\endfoot
    SUN        &  0.49   & 0.030 &   4.82 &  G2V        &  0.00  &  1.6  \\
HD  203244     &  0.47   & 0.030 &   5.42 &  G5V        & -0.21  &    -  \\
HD   17051     &  0.62   & 0.030 &   4.22 &  G0V        & -0.04  &  5.7  \\
HD 1273*       &  0.5    & 0.030 &   5.03 &  G2V \dag    & -0.61  &    -  \\
HD   20407     &  0.41   & 0.042 &   4.82 &  G1V        & -0.55  &    -  \\
HD   20766     &  0.50   & 0.030 &   5.11 &  G2.5V      & -0.25  &    -  \\
HD   20630     &  0.53   & 0.030 &   5.03 &  G5Vvar     &  0.11  &  4.6  \\
HD   20807     &  0.44   & 0.036 &   4.83 &  G1V        & -0.21  &    -  \\
HD   20794     &  0.43   & 0.030 &   5.35 &  G8V        & -0.38  &    -  \\
HD   26491     &  0.50   & 0.058 &   4.54 &  G3V        & -0.23  &    -  \\
HD    1581     &  0.49   &  0.15 &   4.56 &  F9V        & -0.20  &  3    \\
HD   30495     &  0.55   & 0.030 &   4.87 &  G3V        &  0.11  &  3    \\
HD   32778     &  0.42   & 0.030 &   5.28 &  G0V        & -0.61  &    -  \\
HD   34721     &  0.60   & 0.032 &   3.98 &  G0V        & -0.25  &    -  \\
HD   36435     &  0.48   & 0.030 &   5.53 &  G6--G8V    & -0.02  &  4.5  \\
HD   39587     &  0.56   & 0.030 &   4.70 &  G0V        &  0.08  &  9.3  \\
HD   43834     &  0.51   & 0.036 &   5.05 &  G6V        &  0.01  &  1.8  \\
HD   48938     &  0.53   & 0.042 &   4.31 &  G2V        & -0.47  &    -  \\
HD    3443     &  0.47   & 0.036 &   4.61 &  K1V        & -0.16  &  2.7  \\
HD   53705     &  0.60   & 0.036 &   4.51 &  G3V        & -0.30  &    -  \\
HD    3795     &  0.46   & 0.030 &   3.86 &  G3--G5V    & -0.73  &    -  \\
HD   63077     &  0.41   & 0.032 &   4.45 &  G0V        & -0.90  &    -  \\
HD    3823     &  0.60   & 0.042 &   3.86 &  G1V        & -0.35  &  3    \\
HD  64096*     &  0.50   & 0.030 &   4.05 &  G2V \dag    &    -   &    -  \\
HD   65907     &  0.52   & 0.050 &   4.54 &  G0V        & -0.36  &    -  \\
HD   67458     &  0.47   & 0.036 &   4.76 &  G4IV-V     & -0.24  &    -  \\
HD   74772     &  0.89   & 0.032 &  -0.17 &  G5III      & -0.03  &  5.8  \\
HD  202457     &  0.61   & 0.036 &   4.13 &  G5V        & -0.14  &    -  \\
HD  202560     &  0.35   & 0.030 &   8.71 &  M1--M2V \dag&    -   &    -  \\ 
HD  202628     &  0.55   & 0.030 &   4.87 &  G2V        & -0.14  &    -  \\
HD 202940*     &  0.6    & 0.030 &   5.20 &  G5V        & -0.38  &  1.2  \\
HD  211415     &  0.48   & 0.032 &   4.69 &  G3V        & -0.36  &  1.7  \\
HD  211998     &  0.47   & 0.036 &   2.98 &  A3V        & -1.50  &    -  \\ 
HD  212330     &  0.51   & 0.067 &   3.75 &  G3IV       &  0.14  &  1.8  \\
HD  212698     &  0.54   & 0.030 &   4.04 &  G3V        &  0.08  &  9.7  \\
HD   14412     &  0.40   & 0.042 &   5.81 &  G5V        & -0.53  &    -  \\
HD   14802     &  0.62   & 0.032 &   3.48 &  G2V        &  0.10  &  3    \\
HD  104304     &  0.55   & 0.036 &   4.99 &  G9IV       &  0.17  &  1.7  \\
HD  114613     &  0.56   & 0.030 &   3.29 &  G3V \dag    &    -   &  2.7  \\
HD   11695     &  1.02   & 0.030 &  -0.57 &  M4III \dag  &    -   &    -  \\
HD  194640     &  0.46   & 0.032 &   5.17 &  G6--G8V \dag&    -   &    -  \\
HD  20610*     &  0.86   & 0.032 &   0.39 &  K0III      & -0.07  &    -  \\
HD  209100     &  0.39   & 0.030 &   6.89 &  K4.5V      &  0.14  &  0.7  \\ 
HD  211038     &  0.56   & 0.032 &   3.64 &  K0--K1V    &    -   &    -  \\ 
HD  219215     &  1.03   & 0.030 &   0.05 &  M2III \dag  &    -   &    -  \\
HD  29503*     &  0.80   & 0.030 &   1.23 &  K0III      & -0.11  &    -  \\ 
HD   35162     &  0.86   & 0.042 &   0.28 &  G8--K0II--III& -0.31&    -  \\
HD  36079*     &  0.92   & 0.030 &  -0.63 &  G5II       & -0.20  &  5.1  \\
HD    4128     &  0.94   & 0.032 &  -0.30 &  K0III      & -0.01  &  3.3  \\ 
HD   43455     &  1.05   & 0.032 &  -1.55 &  M2.5III \dag&    -   &    -  \\
HD    4398     &  0.82   & 0.032 &   0.44 &  G8--K0III \dag &    -   &    -  \\
HD  102212     &  1.01   & 0.067 &  -0.87 &  M0III \dag  &    -   &    -  \\
HD  111028     &  0.72   & 0.036 &   2.40 &  K1III--IV  & -0.40  &  1.5  \\
HD  112300     &  1.04   & 0.030 &  -0.57 &  M3III      & -0.09  &    -  \\
HD  113226     &  0.94   & 0.030 &   0.37 &  G8IIIvar   &  0.04  &  2.8  \\
HD  114038     &  0.89   & 0.030 &   0.29 &  K1III      & -0.04  &    -  \\
HD  115202     &  0.66   & 0.030 &   2.26 &  K1III \dag  &    -   &    -  \\ 
HD  115659     &  0.92   & 0.030 &  -0.04 &  G8III      & -0.03  &  4.2  \\
HD  117818     &  0.81   & 0.032 &   0.67 &  K0III      & -0.40  &    -  \\
HD 119149*     &  1.15   & 0.030 &  -0.70 &  M2III \dag  &    -   &    -  \\
HD  120477     &  0.90   & 0.032 &  -0.33 &  K5.5IIIvar & -0.23  &  2.2  \\
HD  121299     &  0.83   & 0.032 &   0.70 &  K2III      & -0.03  &    -  \\
HD  123123     &  0.83   & 0.030 &   0.79 &  K2III \dag  & -0.05  &    -  \\ 
HD  124294     &  0.95   & 0.030 &   0.00 &  K2.5IIIb   & -0.45  &    -  \\
HD  125454     &  0.82   & 0.030 &   0.52 &  G8III      & -0.22  &    -  \\
HD  126868     &  0.77   & 0.030 &   1.72 &  G2III \dag  &    -   &  14.3 \\
HD 129312*     &  1.04   & 0.030 &  -1.37 &  G7IIvar    & -0.30  &  6.5  \\
HD  130952     &  0.88   & 0.030 &   0.82 &  G8III      & -0.29  &    -  \\
HD  133165     &  0.83   & 0.032 &   0.64 &  K0.5IIIb   & -0.22  &    -  \\
HD  136514     &  0.85   & 0.030 &   0.98 &  K3IIIvar   & -0.14  &  0.6  \\
HD  138716     &  0.68   & 0.032 &   2.30 &  K1IV       & -0.13  &  2.5  \\ 
HD  140573     &  0.88   & 0.030 &   0.87 &  K2IIIb     &  0.14  &  1.4  \\ 
HD  141680     &  0.82   & 0.032 &   0.68 &  G8III      & -0.28  &  1.1  \\
HD 145001*     &  0.94   & 0.050 &  -0.37 &  G8III      & -0.26  &  9.9  \\
HD 145206*     &  1.00   & 0.032 &  -0.51 &  K4III      &  0.04  &  3.2  \\
HD  146051     &  1.12   & 0.030 &  -0.85 &  M0.5III    &  0.32  &    -  \\
HD  146791     &  0.83   & 0.030 &   0.64 &  G9.5IIIb   & -0.13  &    -  \\ 
HD 148349*     &  0.90   & 0.030 &  -0.71 &  M2 \dag     &    -   &    -  \\
HD 148513*     &  1.10   & 0.036 &  -0.15 &  K4III      & -0.14  &  0.6  \\
HD 150416*     &  1.04   & 0.030 &  -0.48 &  G8II--III  & +0.04  &    -  \\
HD  151217     &  1.00   & 0.032 &   0.00 &  K5IIIvar   & -0.11  &  2.3  \\
HD  152334     &  0.98   & 0.030 &   0.30 &  K4III \dag  &    -   &    -  \\ 
HD  152601     &  0.82   & 0.030 &   0.82 &  K2III      &  0.00  &    -  \\
HD  161096     &  0.91   & 0.030 &   0.76 &  K2III      &  0.05  &  2.7  \\ 
HD 164349*     &  1.17   & 0.030 &  -1.84 &  K0.5IIb    & -0.32  &    -  \\
HD  165760     &  0.91   & 0.095 &   0.33 &  G8III      & -0.15  &  2.2  \\
HD  168723     &  0.67   & 0.030 &   1.84 &  K0III--IV  & -0.10  &  2.6  \\ 
HD 169156*     &  0.77   & 0.030 &   0.82 &  G9IIIb     & -0.17  &    -  \\
HD  169767     &  0.74   & 0.030 &   1.15 &  G8--K0III \dag &    -   &    - \\
HD  170493     &  0.44   & 0.030 &   6.67 &  K3V \dag    &    -   &  3.5  \\ 
HD  171443     &  0.98   & 0.030 &   0.21 &  K3III      &  0.09  &  1.8  \\
HD 171967*     &  1.10   & 0.032 &  -1.57 &  M2III \dag  &    -   &    -  \\
HD 173009*     &  1.11   & 0.032 &  -1.14 &  G8IIb      &  0.05  &  6.0  \\
HD 173764*     &  1.80   & 0.036 &  -2.40 &  G4IIa      & -0.15  &  6.5  \\
HD 175775*     &  1.07   & 0.030 &  -1.76 &  G8--K0II--III& -0.19&    -  \\
HD  176678     &  0.84   & 0.032 &   0.73 &  K1IIIvar   & -0.19  &    -  \\ 
HD  177565     &  0.51   & 0.030 &   4.98 &  G5IV       &  0.03  &    -  \\
HD   17970     &  0.44   & 0.050 &   6.01 &  K1V \dag    &    -   &    -  \\ 
HD 181391*     &  0.77   & 0.030 &   1.61 &  G8III      & -0.21  &  2.8  \\
HD  182572     &  0.60   & 0.032 &   4.27 &  G8IV       &  0.38  &  2.3  \\
HD 183630*     &  1.02   & 0.050 &  -0.90 &  M1IIvar \dag  &    -   &    -  \\
HD  184406     &  0.79   & 0.030 &   1.80 &  K3IIIb     &  0.05  &  1.3  \\ 
HD 186791*     &  1.32   & 0.030 &  -3.02 &  K3II       &  0.00  &  3.5  \\
HD  188310     &  0.91   & 0.032 &   0.73 &  G9IIIb     & -0.32  &  2    \\
HD  189319     &  1.13   & 0.030 &  -1.11 &  K5III \dag  &    -   &    -  \\
HD  190248     &  0.54   & 0.030 &   4.62 &  G7IV       & -0.26  &    -  \\
HD  190406     &  0.49   & 0.030 &   4.56 &  G1V \dag    &    -   &  5    \\
HD  191408     &  0.38   & 0.030 &   6.41 &  K2V \dag    &    -   &  0.1  \\ 
HD  194013     &  0.75   & 0.032 &   0.91 &  G8III--IV  & -0.03  &  1    \\
HD  195135     &  0.88   & 0.036 &   1.07 &  K2III      &  0.03  &    -  \\
HD 196574*     &  0.86   & 0.030 &  -1.04 &  G8III      & -0.13  &  3.7  \\
HD  196758     &  0.89   & 0.036 &   0.77 &  K1III      & -0.12  &  1.8  \\
HD  196761     &  0.42   & 0.030 &   5.53 &  G8--K0V \dag  &    -   &    -  \\
HD 198026*     &  1.03   & 0.030 &  -1.24 &  M3IIIvar \dag &    -   &    -  \\
HD  201381     &  0.73   & 0.032 &   1.00 &  G8III      & -0.15  &  2.8  \\
HD  203504     &  0.86   & 0.030 &   0.71 &  K1III      & -0.14  &  1.2  \\ 
HD  205390     &  0.41   & 0.030 &   6.30 &  K2V \dag    &    -   &    -  \\ 
HD  206067     &  0.87   & 0.032 &   0.76 &  K0III      & -0.17  &  1    \\
HD  206453     &  0.82   & 0.030 &  -0.03 &  G8III      & -0.20  &    -  \\
HD 206778*     &  1.78   & 0.030 &  -4.19 &  K2Ibvar    & -0.05  &  6.5  \\
HD  209747     &  1.06   & 0.030 &   0.32 &  K4III      &  0.00  &  2.3  \\
HD 209750*     &  2.14   & 0.030 &  -3.88 &  G2Ib       &  0.18  &  6.7  \\
HD 211931*     &  0.86   & 0.030 &   1.03 &  A1V \dag    &    -   &    -  \\
HD  212943     &  0.78   & 0.030 &   1.33 &  K0III      & -0.33  &  0.6  \\ 
HD  213042     &  0.44   & 0.030 &   6.71 &  K4V \dag    &    -   &    -  \\ 
HD    2151     &  0.62   & 0.030 &   3.45 &  G2IVvar    & -0.18  &  3    \\
HD 216032*     &  1.03   & 0.030 &  -1.28 &  K5II \dag   &    -   &    -  \\
HD  217357     &  0.33   & 0.030 &   8.33 &  K5--M0V \dag&    -   &    -  \\ 
HD   21749     &  0.38   & 0.030 &   7.01 &  K5V \dag    &    -   &    -  \\ 
HD  217580     &  0.45   & 0.036 &   6.34 &  K4V \dag    &    -   &  3.6  \\
HD  218329     &  1.12   & 0.030 &  -0.43 &  M2III \dag  &    -   &    -  \\
HD  220339     &  0.42   & 0.030 &   6.35 &  K2V \dag    &    -   &  5.5  \\ 
HD  220954     &  0.86   & 0.030 &   0.83 &  K1III      & -0.10  &  0.6  \\ 
HD   27274     &  0.41   & 0.030 &   7.06 &  K5V \dag    &    -   &    -  \\ 
HD   32450     &  0.33   & 0.030 &   8.66 &  M0V \dag    &    -   &    -  \\
HD   42581     &  0.30   & 0.030 &   9.34 &  M1--M2V \dag&    -   &  3    \\ 
HD    4747     &  0.46   & 0.030 &   5.78 &  G8--K0V \dag&    -   &    -  \\
HD  56855*     &  1.76   & 0.030 &  -4.91 &  K3Ib \dag   &    -   &    -  \\
HD  59717*     &  1.05   & 0.030 &  -0.50 &  K5III \dag  &    -   &    -  \\ 
HD  68290*     &  0.84   & 0.030 &   0.95 &  K0III      & -0.03  &    -  \\ 
HD  73840*     &  1.01   & 0.030 &  -0.56 &  K3III      & -0.21  &    -  \\
HD  74918*     &  0.73   & 0.030 &   0.11 &  G8III      & -0.20  &    -  \\
HD   75691     &  0.96   & 0.030 &  -0.01 &  K3III      & -0.11  &    -  \\ 
HD   78647     &  1.65   & 0.030 &  -3.99 &  K4Ib--IIvar&  0.23  &  8.9  \\
HD   81101     &  0.82   & 0.030 &   0.62 &  G6III \dag  &    -   &    -  \\
HD   82668     &  1.21   & 0.361 &  -1.15 &  K5III \dag  &    -   &    -  \\ 
HD   85444     &  0.93   & 0.030 &  -0.50 &  G6--G8III  & -0.14  &  2.9  \\
HD   90432     &  1.04   & 0.030 &  -1.15 &  K4III      & -0.12  &    -  \\ 
HD   93813     &  0.96   & 0.030 &  -0.03 &  K0--K1II   & -0.32  &    -  \\ 
HD   95272     &  0.90   & 0.030 &   0.44 &  K1III      & -0.15  &    -  \\
HD    9540     &  0.50   & 0.030 &   5.52 &  K0V \dag    &    -   &    -  \\ 
HD   98430     &  0.89   & 0.030 &  -0.31 &  K0III      & -0.40  &  1.8  \\ 
\hline
\end{longtable}

%%%%%%%%%%%%%%%%%%%%%%%%%%%%%%%%%%%%%%%%%%%%%%%%%%%%%%%%%%%%%%%%%%%%%%%%%%%%%
\clearpage

\begin{table*}
\begin{center}
\begin{tabular}{c c c c c c }
\hline
\hline
criterion& NUMBER OF STARS &NUMBER OF&\multicolumn{3}{c}{FINAL RESULT}\\
           & REJECTED     & ITERATIONS&$a$& $b$&$\sigma_{WBR}$ \\
\hline
$3\sigma$  &  2   & 2 &  33.2  & -18.0 & 0.60\\
$2.5\sigma$&  5   & 3 &  33.5  & -18.3 & 0.56\\
$2\sigma$  & 16   & 9 &  33.6  & -18.3 & 0.49\\
\hline
 & & &\multicolumn{3}{ l }{ $\sigma_a=0.5\ \ \&\ \   \sigma_b=0.3$. }\\
\hline
\end{tabular}
\end{center}
\caption{The resulting calibrations after three different sigma 
clipping criteria. 
$\sigma_{WBR}$ is the the standard deviation of a single measurement. 
$\sigma_a$ and $\sigma_b$ are the standard deviations 
of the retrieved parameters, respectively, $a$ and $b$ of 
$M_V=a+b\cdot \log W_0$.
The stability of the solution shows the reliability of the WBR as a distance 
indicator.}
\label{sclip}
\end{table*}
%%%%%%%%%%%%%%%%%%%%%%%%%%%%%%%%%%%%%%%%%%%%%%%%%%%%%%%%%%%%%%%%%%%%%%%%%%%%%%

\begin{table}[!ht]
\begin{center}
\begin{tabular}{c c c c}
\hline
\hline
Resolution & $W_0$ [\AA]&$W_{0_{lin corr}}$[\AA]&$W_{0_{quad corr}}$[\AA]\\  
\hline
R=110000 &0.923  & 0.887 & 0.922\\
R=80000  &0.915  & 0.866 & 0.914\\
R=60000  &0.925  & 0.859 & 0.923\\
R=60000  &0.918  & 0.852 & 0.916\\
R=40000  &0.929  & 0.831 & 0.924\\
R=30000  &0.953  & 0.822 & 0.944\\

\hline
Variance & $1.8 \cdot 10^{-4}$&$5.6 \cdot 10^{-4}$&$1.2 \cdot 10^{-4}$\\
\hline
\end{tabular}
\end{center}
\caption{
Six $W_0$ measurements of HD 36079, made on spectra with different
resolutions, both corrected and not corrected for instrumental broadening 
subtracting linearly the projected slit width.
The variance of data in Column 3 data is about 3 times that of Column 1 and,
most important, the corrected $W_0$ decreases with increasing correction. 
These measurements show that the 
linear subtraction of the instrumental profile is not appropriate for our 
data. A quadratic correction seems to be more justified for our data 
(see the last Column).}
\label{resol}
\end{table}

%%%%%%%%%%%%%%%%%%%%%%%%%%%%%%%%%%%%%%%%%%%%%%%%%%%%%%%%%%%%%%%%%%%%%%%%%%%%%%

\begin{table*}[!ht]
\centering
\begin{tabular}{c c c c c c}
\hline
\hline
Column 1  &     Column 2 &     Column 3 &    Column 4  &     Column 5  &     
Column 6  \\
Sanders ID&    $W_0$(\AA)&    m$_V$     &    $M_V(K)$    &    (m-M)$_V$  &
(m-M)$_V$\\
          &              &              &{\tiny no corr}&{\tiny no corr}&
{\tiny corr}\\

\hline  %2         3        4          5         6    
S1010 & 0.851   &  10.48 &  0.657  &   9.823 &  9.742 \\
S1016*& 0.700   &  10.30 &  2.184  &   8.116 &  7.996 \\
S1074*& 0.784   &  10.59 &  1.298  &   9.292 &  9.196 \\
S1135 & 0.963   &   9.37 & -0.310  &   9.680 &  9.617 \\
S1221 & 0.854   &  10.76 &  0.629  &  10.131 & 10.050 \\
S1250*& 0.997   &   9.69 & -0.581  &  10.271 & 10.212 \\
S1479 & 0.868   &  10.55 &  0.502  &  10.048 &  9.970 \\
S1553 & 0.970   &   8.74 & -0.366  &   9.106 &  9.044 \\
S488  & 1.010   &   8.86 & -0.682  &   9.542 &  9.485 \\
S978* & 1.080   &   9.72 & -1.206  &  10.926 & 10.876 \\
\hline
 \multicolumn{4}{r}{ Mean value using all stars: }
                                   & 9.693 & 9.619    \\
 \multicolumn{4}{r}{ Mean value using only unflagged stars: }
                                   & 9.722 & 9.651    \\
\hline 
\end{tabular}
%\end{center}
\caption{Data about the sample of the 10 stars in M67. 
Column 1: Sanders ID number \citep{sanders} of the star.
The stars with doubtful  measurements are flagged with an asterisk.
Column 2: Wilson--Bappu width in \AA.
Column 3: apparent visual magnitude, from Table 1 in \citet{dupree99} 
(see references therein).
Column 4: absolute magnitude inferred from the WBR.
Column 5: retrieved distance modulus.
Column 6: retrieved distance modulus using corrected widths:
$W_{0_{corrected}}=\sqrt{W_0^2-PSW_{R=30000}^2+PSW_{R=60000}^2}$ ($PSW$ is the
projected slit width).}
\label{m67}
\end{table*}

%%%%%%%%%%%%%%%%%%%%%%%%%%%%%%%%%%%%%%%%%%%%%%%%%%%%%%%%%%%%%%%%%%%%%%%%%%%%%%

%-----------------
%-----------------

%-----------------

%-----------------
%-----------------

%-----------------
%-----------------

\end{document}